\documentclass[prd,aps,superscriptaddress,twocolumn,floatfix,nofootinbib]{revtex4}
\pdfoutput=1

\usepackage{amsfonts}
\usepackage{amsmath}
\usepackage{amssymb}
\usepackage{bm}
\usepackage{dcolumn}
\usepackage{graphicx}   
\usepackage[latin1]{inputenc}
\usepackage{latexsym}
\usepackage{rotating}
\usepackage{hyperref}
\usepackage{graphicx}
\usepackage{color}

\newcommand\be{\begin{equation}}
\newcommand\ba{\begin{eqnarray}}
\newcommand\ee{\end{equation}}
\newcommand\ea{\end{eqnarray}}

\begin{document}

\title{Late Time Magnetogenesis from Ultralight Scalar Dark Matter}

\author{Vahid Kamali}
\email{vkamali@ipm.ir}
\affiliation{Department of Physics, McGill University, Montr\'{e}al,
  QC, H3A 2T8, Canada}
\affiliation{Department of Physics, Bu-Ali Sina (Avicenna) University,
Hamedan 65178, 016016, Iran} 
\affiliation{School of Continuing Studies, McGill University, Montr\'{e}al,
QC, H3A 2T5, Canada} 
\affiliation{Trottier Space Institute, Department of Physics, McGill
University, Montr\'{e}al, QC, H3A 2T8, Canada}
  
\author{Robert Brandenberger}
\email{rhb@physics.mcgill.ca}
\affiliation{Department of Physics, McGill University, Montr\'{e}al,
  QC, H3A 2T8, Canada}
\affiliation{Trottier Space Institute, Department of Physics, McGill
University, Montr\'{e}al, QC, H3A 2T8, Canada}


\begin{abstract}

Assuming that Dark Matter is an ultralight scalar field which is coupled to electromagnetism via a gauge-kinetic function and which at the time of recombination is oscillating coherently over a Hubble patch, we show that there is a tachyonic instability for the gauge field modes which leads to the generation of magnetic fields on cosmological scales of sufficient amplitude to explain observations.

\end{abstract}

\maketitle

\section{Introduction} 
\label{sec:intro}

Magnetic fields appear ubiquitous in the cosmos, with  a lower bound of $10^{-17} \mathrm{G}$ ($\mathrm{G}$ stands for Gauss) on the field strength in intergalactic voids at Mpc scales inferred from blazar observations \cite{TaylorVovkNeronov2011, DurrerNeronov2013, Widrow2002, Vachaspati2021}. Traditionally, it has been assumed that such fields must be produced in the early universe. Typically,  early universe proposals generate fields during inflation or phase transitions by breaking conformal invariance in the photon sector \cite{TurnerWidrow1988, Ratra1992, Campanelli2009, CapriniSorbo2014, DurrerHollensteinJain2011, Fujita2015, Adshead2016}. These face well-known challenges (strong coupling, backreaction, and, for phase transitions, correlation-length growth).

A late-time alternative was recently suggested in \cite{Jiao}: an oscillating pseudo-scalar ultralight dark matter field $\phi$ with mass $m$ and with a Chern--Simons (CS) coupling \(g_{\phi\gamma}\phi F\tilde F\) to electromagnetism ($F_{\mu \nu}$ is the field strength of the electromagnetic gauge field $A_{\mu}$ and $g_{\phi\gamma}$ is a coupling constant with dimension of inverse mass) leads to a tachyonic instability for $A_{\mu}$ which starts right after recombination when the electric conductance abruptly falls, producing magnetic fields on all momentum scales $k$ smaller than a critical scale $k_c \propto a\,g_{\phi\gamma}m\phi_0$., where $a$ is the cosmological scale factor normalized to $a = 1$ at the present time, and $\phi_0$ is the amplitude of the oscillations of $\phi$. Most of the energy in fact goes into modes with wavenumber of the order of $k_c$. The resulting amplitude of the magnetic field is limited by backreaction-limited effects. Note that the induced magnetic field in general will have a helical component since the coupling term effects the two helicity modes of $A_{\mu}$ with opposite signs.

In this note we generalize the analysis to the case of an oscillating scalar dark matter field, e.g. a radion or the dilaton, which couples to the gauge kinetic term via a gauge-kinetic function \(I(\phi)\neq 0\).  We find that the tachyonic resonance carries over to this case.  However,  the coupling introduces small oscillations of the fine structure constant \(\alpha\) because \(e_{\rm eff}^2\propto I^{-1}\), and this provides new constraints on the scenario which we discuss.
 
 \section{Framework}

We assume a Friedmann-Lemaitre-Robertson-Walker background spacetime and work in conformal time \(\eta\), primes denoting \(d/d\eta\), and adopt Coulomb gauge \(A_0=0\), \(\nabla\!\cdot\!\bm A=0\). The relevant piece of the action for electromagnetism coupled to a scalar field $\phi$ is
\begin{equation}
\mathcal{L}=\sqrt{-g}\bigg[\frac{M_{\rm Pl}^2}{2}R-\frac12\,\partial_\mu\phi\,\partial^\mu\phi-V(\phi)-\frac14\,I(\phi)\,F_{\mu\nu}F^{\mu\nu}\bigg],\qquad J(\phi)\equiv 0 \, ,
\label{eq:L}
\end{equation}
where $I(\phi)$ is a coupling function. Varying this action with respect to \ \(A_i\) yields the following Fourier mode equation:
\begin{equation}
A_k''+\frac{I'}{I}A_k'+k^2A_k \, = \, 0 \, 
\label{eq:Aeq}
\end{equation}
where the derivative is with respect to conformal time.

Introducing the canonical variable
\begin{equation}
v_k \, \equiv \, \sqrt{I}\,A_k
\end{equation}
removes the ``friction'' term in (\ref{eq:Aeq}) and yields the equation
\be
v_k''+\Big[k^2-\frac{(\sqrt{I})''}{\sqrt{I}}\Big]v_k \, = \, 0
\ee
with
\be
\frac{(\sqrt{I})''}{\sqrt{I}} \, = \, \frac12\frac{I''}{I}-\frac14\Big(\frac{I'}{I}\Big)^2.
\label{eq:veq}
\ee
Equation \eqref{eq:veq} is the equation for a harmonic oscillator with a time-dependent correction to the mass term.  We conclude immediately that there is a tachyonic resonance for sufficiently small values of $k$ provided that the expression $\frac{(\sqrt{I})''}{\sqrt{I}}$ is positive. Resonance occurs for all
\be
k \, < \, k_c
\ee
with
\be
k_c^2 \, = \, \frac{(\sqrt{I})''}{\sqrt{I}} \, .
\ee
Note that, in contrast to the case of a pseudo-scalar field considered in \cite{Jiao}, in the present case the induced magnetic field is not helical since both helicity modes of $A_{mu}$ satisfy the same equation.

If there is a tachyonic resonance, we would expect the time scale of the excitation of $A_{\mu}$ to be much shorter than the Hubble time, in which case we can neglect cosmological expansion, and we can use physical time $t$ instead of conformal time $\eta$.

Note that there are similarities between the dynamics explored here and the dynamics of reheating after inflation. As first pointed out in \cite{TB, DK}, the first stage of the energy transfer between the inflaton (which starts to coherently oscillate about the minimum of its potential after the end of the slow-roll period) and matter proceeds via a parametric resonance of matter fields coupling to the inflaton similar to the one we are exploring here (see e.g. \cite{ABCM, Karouby} for reviews).

\section{Evaluation}

Our scenario is based on the idea that some ultralight dark matter field $\phi$ couples to electromagnetism and induces the coupling term $I$ above (see e.g. \cite{Ferreira, Hui} for reviews on ultralight dark matter).

For concreteness, we take a dilaton-like coupling
\be
I(\phi) \, = \, e^{2\beta\phi/M}
\ee
which results in
\be
\frac{(\sqrt{I})''}{\sqrt{I}} \, = \, \frac{\beta}{M}\phi''+\frac{\beta^2}{M^2}(\phi')^2,
\label{eq:Ichoice}
\ee

We assume that the scalar field $\phi$ is coherently oscillating over the entire Hubble patch at the time of recombination. Such coherent oscillations are induced if there is a misalignment between the early time value of $\phi$ and its late time value. The minimum of the late time effective potential for $\phi$ (which without loss of generality can be taken to be $\phi = 0$) is typically induced by low energy non-perturbative effects.  There is no reason why the early time value of $\phi$ should coincide with $\phi = 0$. For example, in the context of inflationary cosmology our current Hubble patch of space could be emerging from some small fluctuation region during inflation (in the same way that is assumed for the axion, see e.g. \cite{axionrev} for a recent review). Recently, a non-inflationary misalignement mechanism has been proposed \cite{RHBmis} which is based on the assumption that Planck scale effects induce a high temperature effective potential for $\phi$ whose ground state is not $\phi = \phi_0 \neq 0$ (because very different physical effects are involved, this is a rather natural assumption).  In this case, at high temperatures we have $\phi = \phi_0$ uniformly in space.  The value of $\phi$ is then frozen in by Hubble friction until the Hubble expansion rate $H$ drops below the value of the mass of $\phi$.

We can consider two cases for the dynamics of $\phi$ depending on the potential. If $\phi$ evolves like a quintessence field or a dilaton, then we would expect $\phi$ to slowly increase, e.g.  through
\be
\phi^{\prime} \, = \alpha \, ,
\ee
where $\alpha$ is a constant (with dimension of square mass), in which case
\be
\frac{(\sqrt{I})''}{\sqrt{I}} \, = \, \bigl( \frac{\alpha \beta}{M} \bigr)^2
\ee
and we have tachyonic resonance for $k < k_c$ with
\be
k_c \, = \, \frac{\alpha \beta}{M} \, .
\ee
The condition for efficiency of the instability is
\be
k_c \, > \, H(t_{rec}) \, .
\ee
Since the equation of state of $\phi$ in this case corresponds either to kinetion or to dark energy domination,  the case does not apply to $\phi$ as dark matter.  Assuming that the equation of state is that of kination, and that $\phi$ makes up a fraction of order one of the matter energy at the time of recombination, we have $\alpha \sim T_{rec}^2$, and hence the instability criterion reads
\be
\beta \, > \, \frac{M}{m_{pl}} \, ,
\ee
where $m_{pl}$ is the Planck mass.

In the second case (more realistic if $\phi$ is the dark matter field), $\phi$ is oscillating as 
\be
\phi(\eta) \, = \, \Phi {\rm{cos}}(m \eta) \, ,
\ee
where, neglecting the expansion of space and back-reaction effects, the amplitude $\Phi$ is constant. Introducing the constant 
\be
\gamma \, \equiv \, \frac{\beta \Phi}{M} \, 
\ee
we find that to leading order in $\gamma$
\be
\frac{(\sqrt{I})''}{\sqrt{I}} \, = \, - m^2 \gamma {\rm{cos}}(m \eta) \, .
\ee
Thus, we see that for one half of the oscillation period there is a tachyonic driving term for $A_{\mu}$, while for the other half $A_{\mu}$ will oscillate. The net effect is a tachyonic resonance for all values of $k$ with $k < k_c$ with
\be
k_c \, \sim \, \sqrt{\gamma} m \, .
\ee
In contrast to the case of a pseudo-scalar field, in the present case the Floquet exponent $\mu_k$ is independent of $k$. Nevertheless, since the phase space of infrared modes scales as $k^3$, most of the energy is dumped into modes with $k \sim k_c$. The efficiency condition for the resonance in this case becomes
\be \label{eff-cond2}
\beta \, > \, \frac{T_{rec}^2 M}{m_{pl}^2 m} \, .
\ee
 This is a very mild condition as can be seen by assuming $M \sim m_{pl}$, in which case (\ref{eff-cond2}) becomes $\beta > m_{20}^{-1} 10^{-8}$, where $m_{20}$ is the mass of $m$ in units of $10^{-20} {\rm eV}$.
  
\section{Magnetic Field} \label{sec:Bfield}

In this section we estimate the magnetic field induced in the second example of the previous section.

The comoving magnetic power per \(\ln k\) is
\begin{equation}
\mathcal P_B(k,\eta)=\frac{k^5}{2\pi^2 a^4}\,|A_k(\eta)|^2,\qquad A_k=\frac{v_k}{\sqrt{I}} \, .
\label{eq:PB}
\end{equation}
As done in (\cite{Jiao}),  we assume that the resonance and hence the production of $A_{\mu}$ shuts off once a fraction \(F\ll1\) of the energy in $\phi$ has drained into electromagnetism shortly after recombination.  Provided that the resonance is efficient (in the sense discussed in the previous section) the amplitude of power spectrum of the magnetic field at the peak frequency $k_c$ is then given by energy conservation, and is
\be
{\mathcal P}_B(k_c) \, \sim F  \,T_{\rm rec}^4 
\ee
corresponding to a position space amplitude of
\be
B_{\rm rec}(k_c) \, \sim \, F^{1/2} \mathrm{G} \, ,
\label{eq:backreaction}
\ee
where we have used \(1~\mathrm{G} = 1.95\times10^{-20}\,\mathrm{GeV}^2=1.95\times10^{-2}\,\mathrm{eV}^2\). 

Via the standard inverse cascade,  magnetic fields on length scales $\lambda$ larger than $\lambda$ (the inverse of $k_c$) are induced.  The scaling for the non-helical field  is
\be
B(k) \, \simeq \, \Big(\frac{k}{k_c}\Big) B(k_c) \qquad (k<k_c),
\label{eq:slope}
\ee
and thus,  after redshifting \(B\propto a^{-1}\) to today we obtain
\be
\,B_0(k)\ \simeq\ \frac{a_{\rm rec}\,k}{m\sqrt{\gamma}}\ F^{1/2}\ \mathrm{G}\, .
\label{eq:Btoday}
\ee
Evaluating this result on a scale of 1\,Mpc corresponding to the comoving wavenumber \(k_1\simeq 10^{-29}\,\mathrm{eV}\) and making use of \(a_{\rm rec}\simeq 1/1100\), the above equation becomes
\be
B_0(1\,\mathrm{Mpc})\ \simeq\ \frac{10^{-32}\ \mathrm{eV}}{m\sqrt{\gamma}}\ F^{1/2}\ \mathrm{G}.
\label{eq:B1Mpc}
\ee
Inserting the expression for $\gamma$ from the previous section we obtain
\be
B_0(1\,\mathrm{Mpc}) \, \sim \, 10^{-8} \beta^{-1/2} m_{20}^{-1/2} \bigl( \frac{M}{m_{pl}}\bigr)^{1/2} F^{1/2} \mathrm{G} \, .
 \ee
 We see that if $M$ is of the order of the Planck mass $m_{pl}$, and $\beta \sim 1$, then sufficiently large values of the magnetic field can be generated if the mass of $\phi$ is smaller than about $m_{20} \sim 10^{16}$, a very weak constraint.  As the value of $M$ decreases, the bound on $m_{20}$ becomes tighter, but this can be compensated by a decrease in the value of $\beta$.
 
 The bottom line is that there is a large space of parameter space values for which sufficiently large values of the magnetic field on cosmological scales can be generated.
 
\section{Constraints from Time Variation of the Fine Structure Constant}
 
A distinctive feature of the \(I(\phi)F^2\) coupling is an \emph{oscillating} electromagnetic fine structure constant. Canonically normalizing \(A_\mu^{\rm (can)}=\sqrt{I}\,A_\mu\) sends the matter interaction \(e A_\mu J^\mu\) to \(e/\sqrt{I}\,A_\mu^{\rm (can)}J^\mu\),  and hence the effective electromagnetic  fine structure constant is
\begin{equation}
\alpha_{\rm eff}(\eta)=\frac{e_{\rm eff}^2}{4\pi}=\frac{\alpha_0}{I(\phi(\eta))}.
\end{equation}
In the case of the coupling \(I(\phi)=e^{2\beta\phi/M}\) and an oscillating ultralight dark matter field \(\phi=\Phi\cos mt\), the small-amplitude limit yields
\begin{equation}
\frac{\Delta\alpha}{\alpha}\simeq -\,2\gamma\cos mt,
\qquad \gamma=\frac{\beta\Phi}{M}.
\label{eq:alphaAmp}
\end{equation}
Thus,  the same parameter \(\gamma\) that determines the edge \(k_c\) of the tachyonic resonance band also fixes the amplitude of the fractional oscillation of \(\alpha\): \(|\Delta\alpha/\alpha|_{\rm amp}=2\gamma\).

For ultralight dark matter with a quadratic potential energy function, the time-averaged density is \(\rho_\phi=\frac12 m^2\Phi^2\). In a laboratory environment one often takes \(\rho_\phi\simeq \rho_{\rm DM}^{\rm local}\approx 0.3~\mathrm{GeV/cm}^3\), giving
\begin{equation}
\Phi \, = \, \frac{\sqrt{2\rho_\phi}}{m}
\end{equation}
and hence
\begin{equation}
\gamma \, = \, \frac{\beta}{M}\frac{\sqrt{2\rho_\phi}}{m}.
\label{eq:gammaRho}
\end{equation}
Equations \eqref{eq:alphaAmp}--\eqref{eq:gammaRho} permit a direct translation between the clock/Oklo/CMB bounds on \(\Delta\alpha/\alpha\) and the \((m,\beta/M)\) parameter space of magnetogenesis.

The bounds on the time variation of $\alpha$ come from
\begin{itemize}
\item \textbf{Atomic clocks (laboratory, redshift \(z=0\)).} Comparisons of ultra-stable transitions set stringent limits on \emph{oscillatory} \(\Delta\alpha/\alpha\) at frequencies \(\omega\simeq m\) and on slow drift \(\dot\alpha/\alpha\). Such bounds are typically presented as exclusion curves in the phase space of amplitude versus frequency, and translating them uses \(|\Delta\alpha/\alpha|_{\rm amp}=2\gamma\) or, for very slow oscillations, \(|\dot\alpha/\alpha|_{\rm amp}=2\gamma m\). See the reviews \cite{SafronovaRMP2018, Uzan2011, Martins2017}.
\item \textbf{Oklo natural reactor (geophysical, \(z\simeq 0.14\)).} Nuclear resonance analyses constrain \(|\Delta\alpha/\alpha|\) over \(\sim\)Gyr timescales at the \(\lesssim 10^{-8}\) level (see \cite{Uzan2011, Martins2017} for conservative summaries).
\item \textbf{Quasar absorption (astrophysical, \(z\sim 1\!-\!2\)).} Many-multiplet methods deliver ppm-scale constraints on \(\Delta\alpha/\alpha\) with some systematics; see \cite{Uzan2011, Martins2017}.
\item \textbf{CMB (cosmological, \(z\simeq 1100\)).} Recombination physics depends on \(\alpha\) (e.g.\ through the Thomson cross section), yielding \(|\Delta\alpha/\alpha|_{\rm rec}\sim {\cal O}(10^{-3})\) constraints in vanilla models \cite{Uzan2011}.
\end{itemize}
The question now arises as to whether these constraints are consistent with our proposed magnetogenesis mechanism being efficient. 

In order to obey the above constraints, a conservative requirement is
\be \label{gammaupper}
\gamma \, < \, 10^{-8} \, .
\ee
On the other hand, the conditions that the resonance is efficient and that $k_c$ is larger than the cosmological momentum scale $k_{\rm target}$ on which we want to generate a magnetic field lead to a lower bound on $\gamma$ since the requirement is
\begin{equation} \label{conds}
m\sqrt{\gamma} \, \gtrsim\,  \, \max\!\Big[H_{\rm rec},\,k_{\rm target}/a_{\rm rec}\Big] 
\end{equation}
with
\be
k_{\rm target} \, = \, k_1\ \text{for 1\,Mpc},
\label{eq:needs}
\ee
It can easily be checked that both of the conditions in (\ref{conds}) yield a lower bound of the order of
\be \label{gammalower}
\sqrt{\gamma} m \, \geq \, 10^{-26} {\rm eV} \, .
\ee
 It follows that for value of $m$ which are allowed for ultralight dark matter,  namely $m_{20} \gg 10$ \cite{Dalal}, there is a range of values of $\gamma$ for which both bounds (\ref{gammalower}) and (\ref{gammaupper}) can be satisfied.   
 
 The above bound implies an upper bound on the value of the magnetic field which can be obtained on 1 Mpc scales. From (\ref{eq:B1Mpc}) we find
 \be
 B_0(1\,\mathrm{Mpc}) \, < \, 10^{-6} F^{1/2}\ \mathrm{G} \, .
 \ee
Hence, we conclude that there is a large region of parameter space in which our mechanism can generate cosmological magnetic fields of sufficient amplitude to obey the observational lower bound, while at the same time being consistent with the constraints from the varying of the fine structure constant. Taking a less conservative upper bound on $\gamma$ in (\ref{gammaupper}) would further widen the parameter space in which our mechanism can explain the observations.

If \(\alpha\) oscillates during recombination, the ionization history is modulated (e.g.\ via \(\sigma_T\propto\alpha^2\)). A dedicated treatment requires evolving recombination with a small periodic perturbation; however, for \(\gamma\ll 10^{-3}\) the effect is subdominant to current CMB constraints in simple parameterizations \cite{Uzan2011}. This is consistent with the small \(\gamma\) needed for magnetogenesis at late times.

\section{Discussion and Conclusions}

Let us compare our results for scalar field coupling to electromagnetism to the case of the coupling of a pseudoscalar considered in \cite{Jiao}. In the latter setup, we obtain a helical field, while here we obtain a non-helical field. The reason is that in the case of a pseudoscalar coupling, the coupling term enters with opposite sign in the equation of motion for the two polarization modes, while here we obtain a non-helical field since the equation for both polarization modes is the same.  The reason for the difference is thatd a pseudoscalar couples to a pseudoscalar gauge field condensate, while a scalar couples to a scalar condensate. 

In the case of a pseudoscalar field,  the edge of the tachyonic resonance band is given by \(k_c\propto a\,g_{\phi\gamma}m\phi_0\),  the growth exponent by \(\mu_k\propto k^{1/2}\) and the slope of the spectrum of the magnetic field amplitude on large scales given by \(n=3/2\), with a final estimate for the strength of the magnetic field on $1 {\rm Mpc}$ scale given by \(B_0(1\mathrm{\,Mpc})\sim {\tilde{g}}_{\phi\gamma}^{-3/2}F^{1/2}\times10^{-15}\,\mathrm{G}\) after redshifting \cite{Jiao}., where ${\tilde{g}}_{\phi\gamma}$ is the coupling constant between the pseudoscalar field and $F \wedge F$.  In the scalar field case studied here, the edge of the resonance band scales as \(k_c\propto a\,m\sqrt{\gamma}\) and the slope is \(n=1\) (non-helical).  Assuming the same backreaction ansatz \(P_B(k_c)\sim F T_{\rm rec}^4\), which was used in \cite{Jiao} for the pseudoscalar case, it follows from Eq.~\eqref{eq:Btoday} that for natural range of values of \((m,\gamma,F)\) it is possible to obtain a similar amplitude of the magnetic field on Mpc scales as was obtained in \cite{Jiao}. 

The key observational discriminator between scalar and pseudoscalar coupling is the absence or presence, respectively,  of helicity of the induced magnetic field. A further difference are the induced  oscillations of the fine structure constant in the case of scalar dark matter.

In conclusion, we have shown that for ultralight dark matter coupled to electromagnetism there is a tachyonic resonance which can lead to the generation of magnetic fields on cosmological scales. The resonance sets is right after recombination.  There is a large window in parameter space where magnetic fields in excess of $10^{-17} \mathrm{G}$ can be generated.

There are important issues which we have not considered in this work. Firstly, we have assumed that the uniform scalar field condensate persists from the time that the scalar field can start to oscillate (one the Hubble constant drops below the mass) to the time of recombination. We have also neglected the effects of pre-existing magnetic fields at the time of recombination (including those which could be produced by our coupling at earlier times). We have not considered other channels of excitation of inhomogeneous modes of the fields. Finally, we have not computed how back-reaction effects determine the value of the factor $F$.  Most importantly, we have not yet carefully considered the effects of the residual plasma after recombination on the resonance processs. Work on these issues is ongoing.

\section*{Acknowledgement}

\noindent 
The research at McGill is supported in part by funds from NSERC and from the Canada Research Chair program.   VK would like to acknowledge the McGill University Physics Department for hospitality.  RB wishes to thank H. Jiao and J. Fr\"ohlich for collaborating on \cite{Jiao}, and N. Brahma and U-L. Pen for asking challenging questions.


\end{document}